# A random surface theory with non-trivial $\gamma_{string}$

by

J. Ambjørn[1], Z. Burda[2,3,4] , J. Jurkiewicz[1,4] and B. Petersson[2]

## Abstract

We measure by Monte Carlo simulations $\gamma_{string}$ for a model of random surfaces embedded in three dimensional Euclidean space-time. The action of the string is the usual Polyakov action plus an extrinsic curvature term. The system undergoes a phase transition at a finite value $\lambda_c$ of the extrinsic curvature coupling and at the transition point the numerically measured value of $\gamma_{string}(\lambda_c) \approx 0.27 \pm 0.06$. This is consistent with $\gamma_{string}(\lambda_c) = 1/4$, i.e. equal to the first of the non-trivial values of $\gamma_{string}$ between 0 and 1/2.

[1]The Niels Bohr Institute, Blegdamsvej 17, DK-2100 Copenhagen Ø, Denmark
[2]Fakultät für Physik, Universität Bielefeld, Postfach 10 01 31, Bielefeld 33501, Germany
[3]A fellow of the Alexander von Humboldt Foundation
[4]Permanent address: Institute of Physics, Jagellonian University, ul. Reymonta 4, PL-30 059, Krakow 16, Poland

# Introduction

Since the first studies of regularized string theory, i.e. the theory of random surfaces, it has been clear that bounds on the string susceptibility exponent $\gamma$ (also called the entropy exponent) made it difficult to obtain interesting models, i.e. models which have interesting continuum limits when the cut-off is removed. In almost all models considered so far $\gamma \leq 1/2$ is a rigorous bound [1, 2]. In the simplest random surface models like the hyper-cubic lattice model where one sums over all planar random surfaces built of plaquettes of the lattice, it is in addition possible to show that $\gamma > 0$ implies $\gamma = 1/2$. The value $\gamma = 1/2$ is the value for surfaces which are degenerated to so-called branched polymers, and this limit is uninteresting from a string theoretical point of view. If we consider string theories as special examples of matter fields coupled to 2d quantum gravity we have in this broader context many examples of exactly solvable models where $\gamma$ can be calculated analytically. The fractal structure of the random surfaces is a function of the central charge of the conformal field theories coupled to 2d quantum gravity [3]:

$$\gamma(c) = \frac{c - 1 - \sqrt{(1-c)(25-c)}}{12}. \tag{1}$$

However, as is seen from (1) $\gamma \leq 0$ in all cases where the formula makes sense, i.e. for $c \leq 1$. So far there have been no exactly solvable models with $c > 1$ and this includes unfortunately the bosonic string theories of interest since they have $c = d$, the dimension of target space. It has been difficult to obtain genuine random surface models or 2d gravity models coupled to unitary matter theories[5] with $\gamma > 0$. Recently it has been understood that if more than one coupling constant is present, as will be the case for instance if we consider multi-Ising models coupled to gravity, it is possible to obtain a non-trivial behaviour where $\gamma = 1/(n+1)$, $n \geq 2$ [7] (see also [8]). However, the corresponding behaviour will be related to the behaviour of a spin system with central charge $\bar{c} < 1$ (i.e $\bar{c} = 1 - \frac{6}{n(n+1)}$ and $\bar{\gamma} = -1/n$) and

$$\gamma(\bar{c}) = \frac{\bar{\gamma}}{\bar{\gamma} - 1} = \frac{\bar{c} - 1 + \sqrt{(1-\bar{c})(25-\bar{c})}}{12}. \tag{2}$$

The simplest case is $\bar{c} = 0$ and leads to $\gamma(\bar{c}) = 1/3$. This is realized in the case of infinitely many Ising models coupled to gravity [9, 10], and can be considered as a mean field result. However, for a specific multi-spin model it is not known which $\bar{c}$ is selected and this is still a major gap in our understanding of the back-reaction of matter on 2d quantum gravity. Nevertheless, numerical simulations indicate that the back-reaction for relatively small $c$ ($c \leq 4$) is a universal function of $c$, like for $c < 1$, i.e. the fractal geometry of the random surfaces is the same whether we couple eight Ising spins or we couple four q=4 states Potts models to 2d quantum gravity [11].

---

[5]It should be noted that it for some time has been known how to penetrate the barrier $\gamma = 0$ by modifying the concept of random surfaces, such that not only piece-wise linear manifolds but more general simplicial complexes were allowed. In this way one could obtain $\gamma = 1/n$ by using generalizations of the multicritical matrix models, but the corresponding $\bar{c}$ would be less that zero, i.e. non-unitary [4, 5, 6].

If we turn from 2d quantum gravity to the genuine random surface models we can enlarge the coupling constant space by adding an extrinsic curvature term. The motivation for this is that it can be shown that the string tension does not scale in the models without such a term. This is true both for the hyper-cubic model [1] and the model defined by summing over triangulated surfaces [12]. In both cases it is believed that the non-scaling of the string tension is a consequence of the dominance of branched polymers which again might be related to the tachyon problem of the bosonic string. The dominance of branched polymers will diminish if we add an extrinsic curvature term. Further, it can be argued that one gets an extrinsic curvature term by integrating out the fermions of the superstring [13]. For these reasons the models of random surfaces with extrinsic curvature have been extensively studied over the last couple of years in the context of dynamical triangulated random surface models. The present letter reports on the measurement of $\gamma$ at the critical point where the bare string tension and the bare mass in the model scales to zero.

## The model

The model of random surfaces with extrinsic curvature we will discuss has the following partition function [14, 15, 16, 17]:

$$Z(\lambda, A) = \sum_{T \in \mathcal{T}_N} \int \prod_{i=1}^{N} dx_i^\mu \; \delta^{(d)}(\sum x_j^\mu) \; e^{-S_T}, \qquad (3)$$

where $T$ denotes a (closed) triangulation of spherical topology, consisting of of $N$ vertices, $\mathcal{T}_N$ denotes the class of all such triangulations, $x_i^\mu$ the space-time coordinate associated with vertex $i$ of the triangulation, and $\mu$ the space-time index ($\mu = 1, 2, 3$ for $d = 3$). Finally $S_T$ denotes the action corresponding to the particular triangulation:

$$S_T = \frac{1}{2} \sum_{(ij)} (x_i^\mu - x_j^\mu)^2 + \lambda(1 - \cos\theta_{(ij)}). \qquad (4)$$

In this formula $(ij)$ is the link connecting the vertices $i$ and $j$ (if any) and $\theta_{(ij)}$ the angle between the normals to the two oriented triangles $(ijk)$ and $(jil)$ with a common link $(ij)$. This action is the discretized version of the following continuum action:

$$S(\lambda) = \int d^2\tau \sqrt{g} g^{\alpha\beta}(\partial_\alpha x^\mu \partial_\beta x^\mu + \lambda \partial_\alpha n^\mu \partial_\beta n^\mu) + m^2 \int d^2\tau \sqrt{g} \tag{5}$$

The first term is the bosonic string term in the formulation of Polyakov and the $\lambda$-term is a scale invariant extrinsic curvature term. In (5) $g_{\alpha\beta}$ refers to the intrinsic metric. A related version of this action is

$$S(\lambda) = \int d^2\tau \sqrt{h} h^{\alpha\beta}(\mu \, \partial_\alpha x^\mu \partial_\beta x^\mu + \lambda \partial_\alpha n^\mu \partial_\beta n^\mu), \tag{6}$$

where $h_{\alpha\beta}$ refers to the induced metric:

$$h_{\alpha\beta} = \frac{\partial x^\mu}{\partial \tau^\alpha} \frac{\partial x^\mu}{\partial \tau^\beta}, \qquad h^{\alpha\beta} h_{\beta\gamma} = \delta^\alpha_\gamma. \tag{7}$$

Since $\partial_\alpha n^\mu = H_{\alpha\beta} h^{\beta\gamma} \partial_\gamma x^\mu$ (Weingarten eq. for the normals in terms of the second fundamental form $H_{\alpha\beta}$) eq. (6) can be written in a way often used in theory of fluid membranes:

$$S(\lambda) = \int dA(\mu + H^2), \quad H = \left(\frac{1}{r_1} + \frac{1}{r_2}\right), \quad H^2 = H_{\alpha\beta} H_{\gamma\delta} h^{\alpha\gamma} h^{\beta\delta} \tag{8}$$

where $H$ denotes the extrinsic curvature and $r_1(\tau)$ and $r_2(\tau)$ are the principle radii of curvature at the point $x^\mu(\tau_1, \tau_2)$. (5) and (6) are not entirely equivalent at the classical level, but they both produce smoother surfaces with increasing $\lambda$. The action in the form (7) or (8) is not easy to use in the discretized approach since care has to be exercised in order that each physically distinct surface is counted only once. This is automatically satisfied if we use (3) and (4).

Let us briefly summarize what is known about the phase structure of the model. At $\lambda = 0$ we have a regularized version of Polyakov's bosonic string theory in $d = 3$. This theory is known to have a non-vanishing bare string tension [12]. For increasing $\lambda$ we get smoother strings at short distances, but the string still seems to belong to the same universality class as the pure bosonic string. For $\lambda \to \lambda_c \approx 1.45$ we have a genuine phase transition [15, 16, 17, 18] to a string theory where both the mass gap and the string tension seem to scale [17, 19]. It is at this point where we measure the susceptibility exponent $\gamma(\lambda_c)$ which is also called the entropy exponent since it is related to the finite volume partition function. In the discretized approach the internal area of the surface (the equivalent of the continuum $\int d^2\tau \sqrt{g}$) is simply proportional to the number of triangles $N_T$, since each triangle is assumed equilateral in the internal metric. We will take the lattice units such that the area is just

$$\int d^2\tau \sqrt{g} = A_N = N_T = 2N - 4 \qquad (9)$$

where $N$ is the number of vertices of the closed surface of spherical topology. For a fixed number of vertices, or equivalent a fixed internal area A, we have

$$Z(\lambda, A) \sim A^{\gamma(\lambda)-3} e^{\mu_c(\lambda)A} \left(1 + O(\frac{1}{A})\right). \qquad (10)$$

This formula provides us with a convenient way of measuring $\gamma(\lambda)$ by counting so-called baby universes. A minimal neck baby universe, abbreviated "minbu" is a part of the surface connected to the rest of the surface by a minimal loop of links, i.e. a loop of length three. From (9) it follows that the average number of such minbus of area $a < A/2$ for surfaces with the fixed total area $A$ is

$$N_A(a) \sim (a(A-a))^{\gamma(\lambda)-2}. \qquad (11)$$

It is important to note that the leading, non-universal part of the distribution $Z(\lambda, A)$ is not present in $N_A(a)$, and this makes it a convenient observable to use for the determination of $\gamma(\lambda)$. This was first pointed out in [20] and has been verified in practice in a number papers, even in higher–dimensional gravitational theories [21, 22, 11].

# Numerical results

In the simulations of the model we used the vectorization over number of systems proposed in [25] and then developed in [26]. Most of the measurements were performed on lattices of the sizes 800 and 3200 triangles. They were oriented on collecting histograms for the baby universe distribution. To estimate the autocorrelation time we additionally measured the radius of gyration which is known to be the slowest mode in the algorithm. On the smaller lattice we simulated 64 systems in parallel, while for the larger, because of the memory limitations we had to lower their number to 32. The measurements were taken every 100 sweeps. The measurement of the minbu distribution is not vectorizable, so we ran it in the scalar mode. The maximal size of the lattice in our simulation was not only limited by the memory but also by the autocorrelation time $\tau$ which is particularly high in the critical region where we spent most of the computer time. From the measurements of the gyration radius we estimated $\tau$. In the critical region $\tau$ is $0.13(4) \cdot 10^6$ sweeps for the 3200 triangle lattice and $0.14(3) \cdot 10^5$ sweeps for the 800 triangle one. For the larger lattice we run $3.0 \cdot 10^6$ sweeps in each of 32 systems while for the smaller one $1.0 \cdot 10^6$ sweeps in each of 64 systems. This corresponds around to 23, 70 autocorrelation times, respectively. At the beginning we made $10^6$ and $10^5$ sweeps, respectively, for thermalization, which in both cases corresponds to $7\tau$. One can worry about insufficient thermalization. We checked, however, that discarding the data from the next few autocorrelation times does not change the the resulting histograms for minbu distribution. In fact, the autocorrelation times for the quantities describing the intrinsic geometry are much lower for the studied range of lattice sizes than for the gyration radius [23].

We simulated systems for different values of the coupling $\lambda$. The resulting minbu distribution can be divided into three classes corresponding to two different phases of the model and to the critical region. In fig.1 we show typical curves for each of these cases for the system with 3200 triangles. They are plotted in log log scale. Going downwards the curves correspond to $\lambda = 0.5, 1.45, 1.9$.

The slopes of the curves in the crumpled phase and in the critical region differ slightly. Both are close to $-2$. As one can see from the formula (11) the slopes are dominated by the gamma independent power $-2$ in the minbu distribution. To visualize the $\gamma$–dependent information in the minbu distribution, we plot in fig. 2 the function $F(a) = N_A(a)(a(1 - a/A))^2$ which explicitly removes the bulk power $-2$. The function $F(a)$ is plotted in log log scale against $a(1-a/A)$. It is clearly seen that the upper curve, $\lambda = 0.5$ has a negative slope, although it's value changes along the curve. This is a well known problem for $2d$ gravity models coupled to purely gaussian fields or multi–Ising spins with $c \geq 1$ [11] and might be due to logarithmic corrections. We obtained this type of curves for small $\lambda$'s which correspond to being deep in the crumpled phase, and it confirms that the model is dominated by the gaussian term and belongs to the same universality class for this range of $\lambda$. The gaussian model itself was discussed in more detail elsewhere [11].

When approaching the critical region the curves gradually change until in the critical region their slope becomes positive. The situation is in this respect very similar to that observed in the spin models where the effective numerical value of $\gamma$, as a result of the finite size effects, passes gradually from that for pure gravity outside the critical region, to the value characteristic for critical matter sector, in the vicinity of the pseudocritical value of the coupling constant. In our case the critical region, where the slope stayed roughly constant corresponds to $\lambda$ between 1.4 and 1.5. As values of $\lambda$ for which we performed the high statistics simulation we chose $\lambda = 1.44$ for 800 triangles and $\lambda = 1.45$ for 3200. The choice of the pseudocritical values of the coupling constant was dictated by estimation of positions of the heat capacity maxima from the results for the smaller systems and is slightly higher than the value $\lambda_c \approx 1.425$ quoted in [18]. As was explained above the small difference has no effect on the measured value of $\gamma$.

| $N_t$ | $\gamma$ |
|-------|----------|
| 800   | 0.31(7)  |
| 3200  | 0.27(6)  |

Table 1:

The lowest curve in fig.1 corresponds to the 'flat' phase where the normals become almost aligned. In this phase the acceptance rate goes down below 10% and the autocorrelation time $\tau$ becomes even larger. The phase seems to be a lattice artifact and one would need much larger lattices and probably a new algorithm in this phase.

To extract $\gamma$ we fitted our data points directly to the formula (11). As input to our fits we took the data points $N_A(a)$ averaged over all systems. Their errors $\delta N_A(a)$ were naturally provided by the widths of the distributions in different systems. As is clearly seen from the figure 2, both a quality of the fit and the value of $\gamma$ depend on the range of the minbu areas $a$ used for the fit. We made different fits for different ranges $a_{min} \leq a \leq a_{max}$. The values of $\gamma$ are plotted for different values of the lower limit $a_{min}$ in fig.3 and fig.4. Different symbols correspond to different values of the upper limit $a_{max}$. The fitted values of $\gamma$ saturate with increasing $a_{min}$. The corresponding values of $\chi^2$ decrease. At $a_{min}$ around 30, $\chi^2$ becomes of the order one per degree of freedom and from there on it stays at this level only weakly depending on $a_{max}$. This means that when one excludes the minbus of the size smaller than 30 one gets surprisingly good fit to the formula $N_A(a)$ (11) in the broad range of $a$ without including the next-to-leading terms which were necessary, for example, in the pure gravity model. Although the values of $\gamma$ look very stable as a function of the limits $a_{min}$ and $a_{max}$ one can observe a tendency that they seem to decrease when the limits are increased. This is seen in fig.4 for the 3200 triangle lattice, where the effective $\gamma$ goes gradually down after the maximum around $a_{min} = 30$. The values decreased also with increasing $a_{max}$. The same tendency is also reflected in the fact that the values of $\gamma$ are lower for the larger lattice. All this can be summarized by saying that the effective value of $\gamma$ decreases for larger areas. The results for the effective $\gamma$ are summarized in the table 1.

# Discussion

The measurements performed in this paper were done for systems with finite size. The measured values of $\gamma_c$ indicate that $1/4 < \gamma_c < 1/3$. The error bars quoted are probably over–estimated. For increasing size we see a clear trend to lower the value of $\gamma_c$ which suggests that for the infinite system we have $\gamma(\lambda) \approx 1/4$ for the ensemble of random surfaces described by the partition function (3) at $\lambda \approx \lambda_c$. As explained in the introduction this implies that the surface theory given by (3) realizes the first non-trivial exponent $\gamma = 1/4$ of the possible values $\gamma = 1/n$, $n \geq 2$. The corresponding "effective" central charge is $\bar{c} = 1/2$ and it indicates an equivalent fermionic description of the random surfaces with extrinsic curvature. It is worth to recall that the same observation has been made a long time ago for hyper-cubic surfaces in d=4: Extensive computer simulations [24] showed that $\gamma \approx 1/4$ for a model where "back-tracking" of plaquettes[6] was not allowed. On the other hand $\gamma \approx 1/2$ for d=8 in the hyper-cubic lattice model without back-tracking. This suggests that at least for low dimensions we have a universal non-trivial susceptibility exponent for random surfaces with extrinsic curvature. It should be most interesting to check at which dimension the exponent $\gamma$ of dynamically triangulated surfaces changes from 1/4 to 1/2, and to get an understanding of this from the point of view of continuum string or field theory.

# Acknowledgments

The computation was performed on the CRAY-YMP at HLRZ in Jülich, which we would like to acknowledge here. One of us (Z.B.) wishes to thank Alexander von Humboldt Foundation for the fellowship. This work was partially supported by the KBN grant 2P30204705.

---

[6] By back-tracking we mean that two neighbouring plaquettes occupy the same physical space on the lattice. Although this is not allowed in the model it *does* allow non-neighbouring plaquettes to do so, which means that it is only a local constraint. On the hyper-cubic lattice the neighbouring plaquettes can have only a discrete number of angles relative to each other, and in the case of extrinsic curvature the statistical weight of the surface will depend on these angles, suppressing surfaces where many plaquettes have large angles. To forbid back-tracking corresponds to taking the coupling constant for a bending of 180 degrees to be infinite.

# Figure captions

Fig.1 The minbu distribution $N_A(a)$ in log log scale, for the lattice with $A = 3200$ triangles, for the different values of the coupling $\lambda$ : 0.5 (+), 1.45 ($\diamond$), and 1.9 ($\square$)

Fig.2 The function $F(A) = N_A(a) \cdot (a(a - 1/A))^2$ versus $a$ in log log scale for the lattice with $A = 3200$ triangles.

Fig.3 The effective values of $\gamma$ for the lattice with $A = 800$ triangles, for different fit ranges : $a_{min}$ runs along the vertical axis, while different $a_{max}$ are represented by : 200 ($\diamond$), 400 (+), 600 ($\square$) and 800 ($\times$).

Fig.4 The effective values of $\gamma$ for the lattice with $A = 3200$ triangles, for different fit ranges : $a_{min}$ runs along the vertical axis, while different $a_{max}$ are represented by : 150 ($\diamond$), 200 (+), 250 ($\square$) and 300 ($\times$).

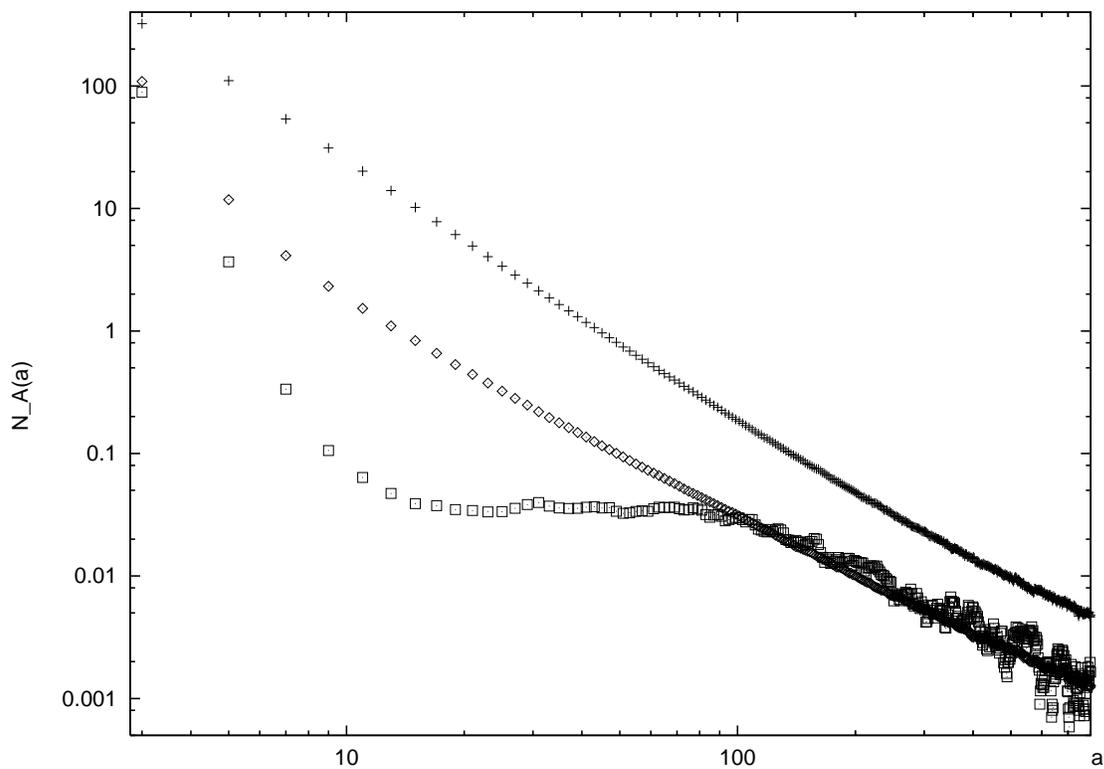

Figure 1

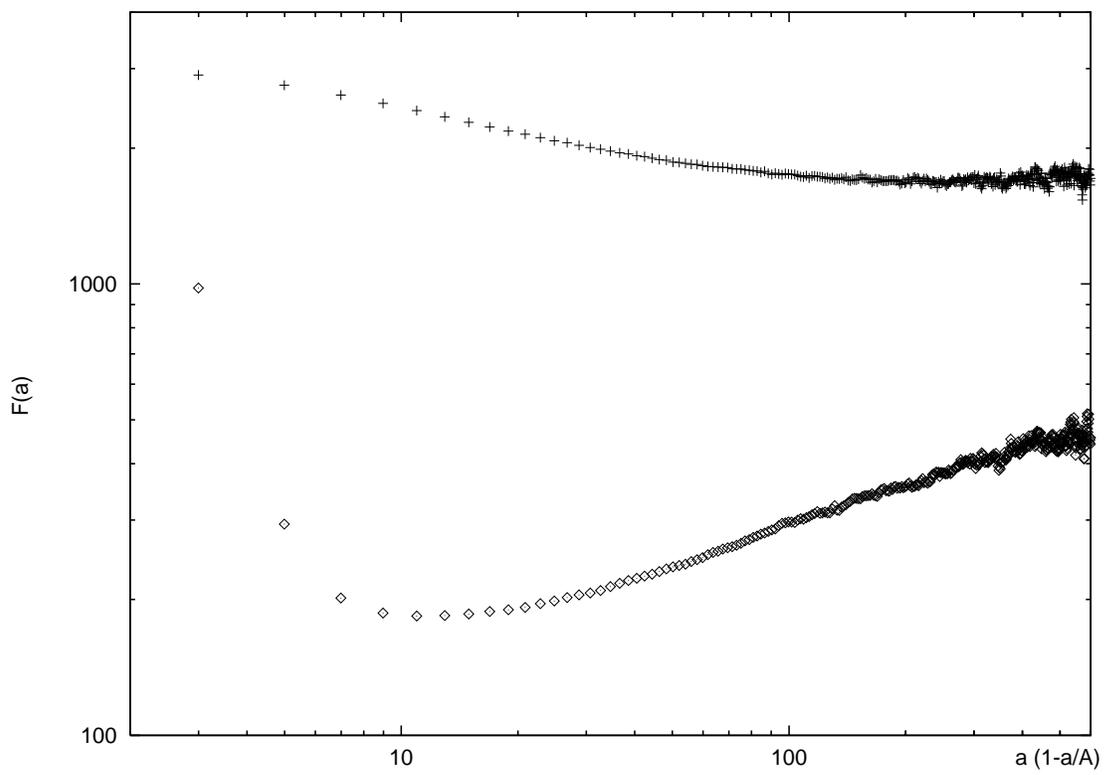

Figure 2

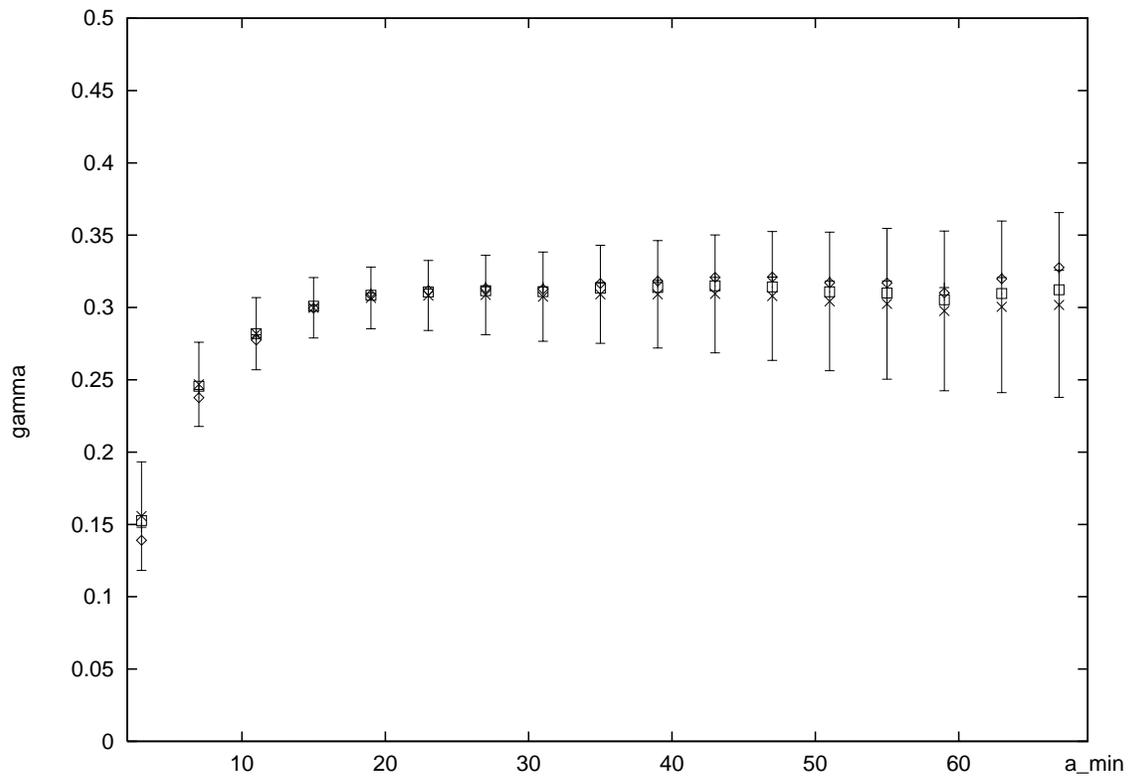

Figure 3

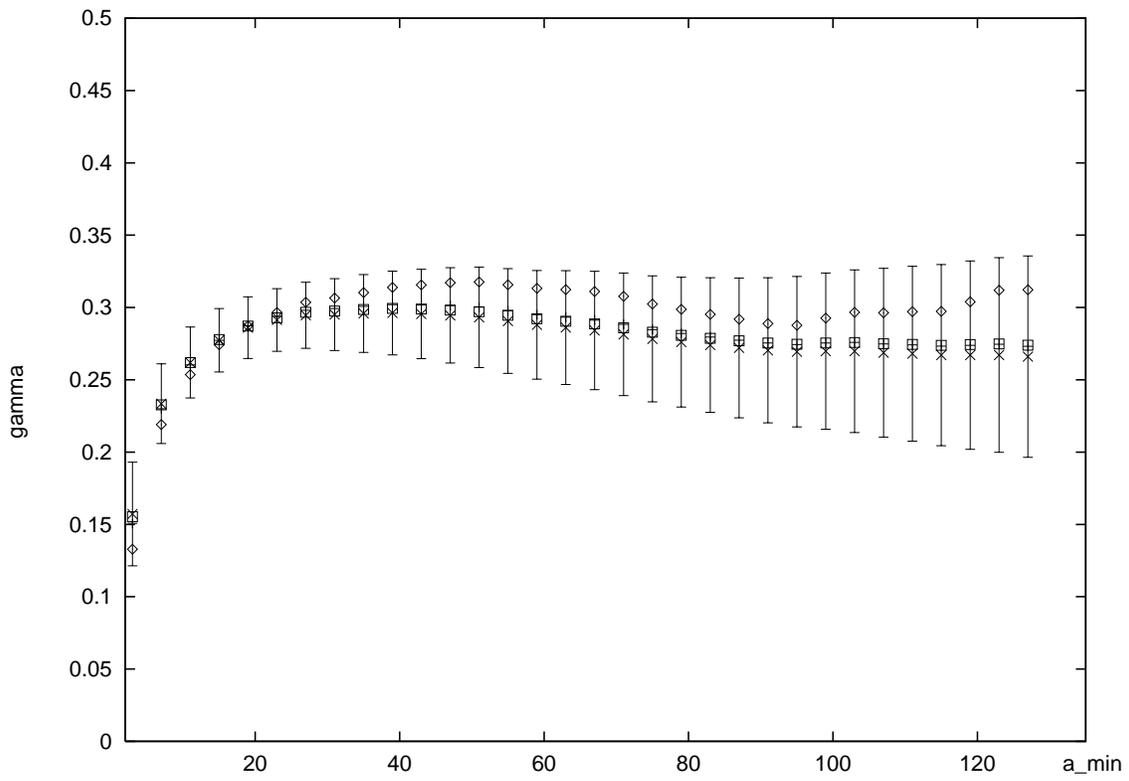

Figure 4